\documentclass[aps,prb,twocolumn,floatfix,%
showpacs,showkeys,amssymb]{revtex4}
\usepackage{graphicx}

\begin{document}

\date{\today} 

\def\ba{\begin{array}}
\def\ea{\end{array}}
\def\be{\begin{equation}\begin{array}{l}}
\def\ee{\end{array}\end{equation}}
\def\bea{\begin{equation}\begin{array}{l}}
\def\eea{\end{array}\end{equation}}
\def\f#1#2{\frac{\displaystyle #1}{\displaystyle #2}}
\def\om{\omega}
\def\omm{\omega^a_b}
\def\we{\wedge}
\def\de{\delta}
\def\De{\Delta}
\def\va{\varepsilon}
\def\omb{\bar{\omega}}
\def\la{\lambda}
\def\vv{\f{V}{\la^d}}
\def\si{\sigma}
\def\t{T_+}
\def\v{v_{cl}}
\def\m{m_{cl}}
\def\n{N_{cl}}
\def\bi{\bibitem}
\def\c{\cite}
\def\sa{\sigma_{\alpha}}
\def\ua{\uparrow}
\def\da{\downarrow}
\def\mua{\mu_{\alpha}}
\def\ga{\gamma_{\alpha}}
\def\g{\gamma}
\def\ora{\overrightarrow}
\def\pa{\partial}
\def\ov{\ora{v}}
\def\al{\alpha}
\def\bt{\beta}
\def\R{R_{eff}}
\def\th{\theta}

\def\muu{\f{\mu}{ed}}
\def\E{\f{edE(\tau)}{\om}}
\def\t{\tau}

\title{Polarons with a twist}

\author{Wei Zhang, Alexander O. Govorov, and Sergio E. Ulloa}

\affiliation{Department of Physics and Astronomy, and Nanoscale
and Quantum Phenomena Institute, Ohio University, Athens, Ohio
45701-2979}

\begin{abstract}
We consider a polaron model where molecular \emph{rotations} are
important. Here, the usual hopping between neighboring sites is
affected directly by the electron-phonon interaction via a {\em
twist-dependent} hopping amplitude. This model may be of relevance
for electronic transport in complex molecules and polymers with
torsional degrees of freedom, such as DNA, as well as in molecular
electronics experiments where molecular twist motion is
significant. We use a tight-binding representation and find that
very different polaronic properties are already exhibited by a
two-site model -- these are due to the nonlinearity of the
restoring force of the twist excitations, and of the
electron-phonon interaction in the model. In the adiabatic regime,
where electrons move in a {\em low}-frequency field of
twisting-phonons, the effective splitting of the energy levels
increases with coupling strength. The bandwidth in a long chain
shows a power-law suppression with coupling, unlike the typical
exponential dependence due to linear phonons.
\end{abstract}


\maketitle

There has been considerable interest in the study of electronic
transport in organic materials.  They provide technologically
relevant alternatives for new electronic and optical materials,
and offer an opportunity to gain deeper understanding of quantum
transport processes. Moreover, transport in molecules also
constitutes the basis for important biological phenomena, such as
photosynthesis and other charge transfer events. \c{book1} Recent
developments in the fabrication of high electron mobility systems
using organic semiconductors, have further emphasized the need to
understand mobile polaronic behavior in complex molecular
crystals. \c{schon} In addition, a series of beautiful experiments
in the area of molecular electronics have underscored the
importance of atomic vibrations in the measured current-voltage
characteristics of these systems. \c{Reed, Reifenberger, Zhitenev,
McEuen}

In typical inorganic materials the atomic degrees of freedom
(\textbf{DOF}) associated with \textit{translation} are the most
important in electronic transport.  In contrast, the
\textit{twisting} DOF (or ``librons") of molecular complexes may
also play important roles in organic and molecular solids. In a
wide variety of organic linear-chain crystals of stacked
molecules, the librons may in fact be essential.  For instance, it
was found that libron scattering was important in the
understanding of conductivity mechanisms in the TTF-TCNQ system,
\c{ttf1,ttf2,ttf3,ttf4} as well as in phthalocyanine-based
``molecular metals". \c{hale} More recently, the conductivity of
DNA and constituent base  crystals has been studied
experimentally, \c{expa,expb,expc,expd,expe} and theoretically.
\c{th1,yu,bru,la,C00} Since DNA can be viewed as a ``molecular
wire", a one-dimensional twisted chain of stacked base pairs with
a somewhat flexible structure, \cite{B00} it is likely that
parallel rotation of adjacent base pairs will be important in
electronic transfer mechanisms.  Notice that these modes would
have a stronger impact on the electronic overlaps along the chain
than other flexural modes. In fact, a variable range hopping
theory where the twist angle fluctuations between base pairs are
taken into account was used recently to explain the temperature
dependence of electrical conductivity along the DNA double helix.
\c{yu} Bruinsma {\em et al.} have also studied charge migration
along DNA based on a model in which large structure fluctuations,
including relative rotation between adjacent bases, are
considered. \c{bru}

There exist a few models of electron-libron interactions in the
literature. Many of them have considered the situation of small
rotation angles in a harmonic or nearly harmonic approximation.
\c{ttf3,ttf4,hale,bru,B00}  Although other authors have considered
large angle rotations, \c{yu,bar} the molecular kinetic energy
term was neglected, and thus the inertial backreaction effects in
molecules  were not taken into account. Our model considers the
possibility of large rotation effects, which are not negligible in
systems such as (TTF)$_7$I$_5$, and others with similarly flexible
structure. We also include the full kinetics of the molecule in
order to include backreaction effects. Our description is fully
quantum mechanical and includes strong anharmonic interactions,
shown to be essential for polaron self-trapping. \c{RBK} From the
full solution of a two-site model, we find that in the adiabatic
regime, where electrons move in a {\em low} frequency field of
twisting-phonons, the splitting of the energy levels (and
therefore the effective intersite hopping) increases with phonon
coupling strength. In the opposite, non-adiabatic regime, the
effective hopping constant increases with phonon frequency, but
saturates at high values.   For a long chain model with
local-twists, we find a power-law suppression of the bandwidth,
unlike the exponential result for translational polarons.  These
results arise from the different coupling scheme and the nonlinear
twist-polaron restoring force, as we now
describe. \\

 \noindent {\bf The twist polaron model}.
The on-site interaction of the electron with phonon modes tends to
localize the electron. On the other hand, the molecular transfer
(hopping) integral in a given atomic or molecular configuration
results in electron delocalization. In an extended system, the
hopping integral competes with the phonon interaction, and
depending on the relative strengths of the different terms, one
obtains a mobile polaron (with a heavier effective mass), or a
self-localized polaron. \c{fro,holst}  Our model allows for such
physical behavior as well. Most importantly, however, the phonon
enters here modifying also the hopping integral and is then
substantially different to typical models for translation-phonon
polarons. \c{pol-book}

For simplicity, let us first consider a two-site model. The
dynamics of the system is described by the Hamiltonian
 \begin{eqnarray} H &=&-(c_1^+c_1+c_2^+c_2)\va^*\cos \theta \\ \nonumber
  & & - (t_0+t_1\cos \th )(c_1^+c_2+c_2^+c_1) \\ \nonumber
 & &-\f{\hbar^2}{2I}\f{\pa^2}{\pa \th^2}- K\cos \th \, ,
 \end{eqnarray}
where the first term represents the possible on-site energy
modulation due to the molecular rotations with angular DOF $\th$,
and the $c_i^+$ operators describe the two electronic wave
functions (``sites"). The second term is the rotation-dependent
hopping term. Notice that the molecular rotation would either
enhance or suppress electron hopping in this model, depending on
whether $t_1$ and $t_0$  have the same or opposite sign. A
microscopic calculation of stacked molecules, such as the base
pairs in DNA, or the phthalocyanine stacks, allows one to estimate
the effect of molecular twists on the electron hopping or overlap.
\cite{B00} Recent results for guanosine crystals, \c{moli} for
example, show clearly that eclipsing  of molecules in successive
``layers" is essential for substantial electron hopping between
them.  This twist dependence is introduced in our model via the
hopping amplitude $t_1$, and then plays the role of a coupling
constant. The last two terms in (1) are the relative kinetic
energy for the molecules and the interaction potential
characterized by the libron restoring `force constant' $K$. Here,
$\th=\th_1-\th_2$, where $\th_i$ is the rotation angle of each
molecule, and $I$ is the reduced moment of inertia for the
relative rotation of the two adjacent molecules (see inset Fig.\
1). An overall rotation of the pair is gapless and neglected.
Notice also that the first term in (1) leads to a constant shift
of $K$, which is then implicit in the following discussion. Notice
that this model considers nonlinear coupling to the electrons via
$t_1$, \c{RBK} as well as a non-linear restoring force for the
relative rotation of this two-molecule stack. On the other hand,
the model does not consider the possible mixing of rotation DOF
with translation displacements, although this coupling is for the
most part small. \c{bru}  We believe that our model contains the
main physics of polaronic excitations corresponding to the
rotation DOF.

In the following, we use the energy unit ${\hbar^2}/{2I}$ for
convenience, so that all energy quantities are dimensionless from
now on. In order to solve the eigenvalue problem $H\psi=E\psi$, we
use
 \be
 | \psi \rangle=\delta(\th)c^+_1|0 \rangle+\bt(\th)c^+_2 |0 \rangle \, ,
 \ee
which yields, from $(H-\va 1)\psi=0$, a set of secular equations
for $\delta$ and $\beta$.  Defining $A=\delta+\bt$, and
$B=\delta-\bt$, we have
 \bea
 (\f{\pa^2}{\pa\th^2}+\va+t_0+p\cos\th)A =0\\
 (\f{\pa^2}{\pa\th^2}+\va-t_0+q\cos\th)B =0 \, ,
 \eea
 with $q=K-t_1$, and $p=K+t_1$. The solutions are periodic Mathieu
functions, \c{math} with properties which depend on the values of
the coefficients $p$ and $q$.

As we will see below, the rotation DOF in the model produce
modifications of the electronic hopping. In order to reveal the
physics more clearly, we analyze the two-site problem in more
detail in the adiabatic and non-adiabatic regimes. \\

 \noindent {\bf Non-adiabatic regime}.
In the non-adiabatic regime $K\gg t_0,t_1$, the electrons move in
a field of high frequency twist-phonons. We first neglect fully
the hopping terms. We find two degenerate states
 \bea
 \psi_1=\left(\ba{c} \phi_0\\0 \ea \right)\, , \,\,\,
 \psi_2=\left(\ba{c} 0\\\phi_0 \ea \right)\, ,
 \eea
where $\phi_0$ is the normalized ground state (with eigenvalue
$\va_0$), solution of the equation
 \be
 \f{d^2\phi}{d\th^2}+[\va+K\cos\th]\phi(\th)=0 \, .
 \ee

To calculate the first order correction, we take the wave function
$\psi=\al \psi_1+\beta \psi_2$, so that the secular equation
becomes
 \bea
 \left(\ba{cc} \va_0-\va & -(t_0+t'_1) \\
 -(t_0+t'_1)& \va_0-\va \ea \right) \left(\ba{c} \al \\ \beta\ea
 \right)=0 \, ,
 \eea
where the renormalized twist-dependent hopping constant is
 \begin{equation}
 t'_1=t_1 \int \phi_0^* \,  \cos\th \, \phi_0 \, d\th \, .
 \label{t1p}
 \end{equation}
The eigenvalues are simply $\va_{1,2}=\va_0\pm |t_0+t'_1|$, so
that the splitting of the two originally degenerate eigenvalues is
$\De\va=2|t_0+t'_1|$. It is easy to see from Eq.\ (\ref{t1p}) that
$t'_1\leq t_1$. Therefore, $\De\va\leq T$, where $T=2t_0+2t_1$ is
the hopping constant ``without rotation", i.e. $\theta \equiv 0$.

In the regime of large $K$ ($K\gg 1$), the local twist potential
is effectively harmonic and one can write the solution as a
superposition of local harmonic oscillators about $2m\pi$,
 \begin{equation}
 \phi_0=\sum_m
 \frac{\al^{1/2}}{\pi^{1/4}}e^{-\al^2(\th-2m\pi)^2/2} \, ,
 \end{equation}
where $\alpha ^4 = K/2$, and the eigenvalue is $\va_0=-K+ \alpha
^2$. Correspondingly,
 \begin{equation}
  t'_1=t_1 e^{-1/4\al^2} \simeq t_1 \left(1-\frac{1}{4\al^2} \right) \, ,
 \end{equation}
and $\De \va=2t_0+2t_1(1-1/4\al^2) \lesssim T$, with $\De\va$
approaching $T$ in the large $K$ limit.  In this case, the
molecule is weakly oscillating about $\th = 0$ (mod $2 \pi$), with
a relatively minor impact on the electron hopping.  Notice that
the resulting ``bandwidth narrowing" is similar {\em in form} to
the result of a two-site vibrational polaron. \c{pol-book} There,
the effective bandwidth in the non-adiabatic regime is $\tilde{J}
= J e^{-\gamma^2/2}$, where $J$ is the original bandwidth and
$\gamma$ is the electron-phonon coupling constant.  Notice however
the different role that $K$ plays here, and that the bandwidth is
only weakly suppressed in this case.

When $K$ is small ($1\gg K \gg t_0,t_1$), the ground state
function can be written as, \c{math}
 \begin{equation}
 \phi_0(\th) \simeq \frac{1}{\sqrt{2\pi(1+K^2/2)}}(1+K\cos\th) \, ,
 \end{equation}
and the eigenvalue is $\va_0=-\f{1}{2}K^2$. We find that $t'_1=t_1
\f{K}{1+K^2/2} \simeq Kt_1$, and $\De \va=2t_0+2Kt_1 \ll T$.
Notice that in this limit of \textit{soft} restoring force, the
molecules are nearly uniformly rotating at all angles, and
strongly suppress the electron intersite hopping.

We see that with increasing $K$, the renormalized twist-dependent
hopping constant $t'_1$ increases and saturates to $t_1$, as shown
schematically in Fig.\ 1. We should emphasize that one can view
the resulting $\De \va$ as akin to a bandwidth in the long chain
limit. These results then indicate that the twisting DOF strongly
prevent the electron from propagating along the chain. The
``stiffer" the rotation mode or libron, the easier for the
electrons to move from site to site. \\

 \noindent {\bf Adiabatic regime}.
In this limit, $K\ll t_0,t_1$, the electron moves `rapidly' in a
field of low frequency `slow' phonons. We assume a wave function
of the form
 \bea
 \left(\ba{c} \al(\th) \\ \beta (\th)\ea\right) =S(\th)
 \left(\ba{c} s_1(\th) \\ s_2(\th)\ea\right) \, ,
 \eea
and first consider the electron moving in a frozen background,
 \bea
 \left(\ba{cc} -E_0(\th) & -(t_0+t_1\cos\th) \\
 -(t_0+t_1\cos\th) & -E_0(\th) \ea\right) \left(\ba{c}
 s_1(\th) \\ s_2(\th) \ea\right) =0 \, .
 \eea
The corresponding eigenvalues $E_0^{\pm}=\pm|t_0+t_1\cos\th |$
play the role of a potential energy term in the equations for
$S(\th)$,
 \begin{equation}
 \left \{\f{\pa^2}{\pa\th^2}+K \cos\th-E_0^{\pm}(\th)+\va\right\}
S(\th)=0 \, .
 \end{equation}
These equations can also be solved in terms of Mathieu functions.
It is easy to find that when $K\ll t_0,t_1$, and $t_1\gg 1$, the
eigenvalues of the problem are given by
 \begin{eqnarray}
 \va_1 &=& \va_{op} - t_0 -K +\f{K}{\sqrt{8t_1}} \\ \nonumber
 \va_2 &=& \va_{op} +t_0 +K -\f{K}{\sqrt{8t_1}} \, ,
 \end{eqnarray}
where $\va_{op} =-t_1+\sqrt{t_1/2}$ is the polaronic energy shift
due to the electron-libron off-site interaction. In this case,
$\De\va=2t_0 +2K-{K}/{\sqrt{2t_1}}$.  Unlike the non-adiabatic
case, $\De\va$ saturates to $2t_0+2K$ for increasing $t_1$, and it
therefore shows a strong bandwidth narrowing.  Notice, however,
that the two-site adiabatic problem for the small Holstein polaron
yields an even stronger suppression of tunneling.  In that case,
\c{pol-book} one finds an exponential narrowing with coupling,
$\sim e^{-\gamma^2/2}$, similar to the non-adiabatic case
mentioned above, but with large $\gamma$ values. \\

 \noindent {\bf Long chain rotational polaron}.
The long-chain generalization of the two-site model is an
interesting physical problem with additional mathematical
complexity which will be reported elsewhere. \cite{toberep} Here,
we focus on a simpler local-twist model involving local angular
coordinates.   The local-twist hamiltonian is given by
\begin{widetext}
 \begin{equation}
 H = \sum_i \left \{ -\frac{\hbar^2}{2I} \frac{\partial ^2}{\partial \theta
 _i ^2} - K \cos \theta _i - \va _1 c^+_i c_i \cos \theta _i -
 t_0 (c^+_i c_{i+1} + c^+_{i+1} c_i ) \right \} \, ,
 \end{equation}
\end{widetext}
which includes anharmonic restoring force and local non-linear
electron-phonon coupling via $\va _1$ (the more general
$t_1$-dependence is also discussed in Ref.\ \onlinecite{toberep}).

In the non-adiabatic regime, $K \gg 1$, one can generalize the
two-site approach and identify two limiting cases.  For strong
restoring force, $K \gg \va_1$, it is possible to show that the
polaron energy shift is $- \va_1$, while the bandwidth is slightly
reduced to $\Delta = 2 t_0 (1-\va_1 /K)$.  In the case of softer
twist frequency, $\va_1 \gg K \gg 1$, the bandwidth is given by
$2t_0 (K/\va_1)^{1/8}$.  This power law suppression of hopping is
much weaker than the usual Holstein `translational' polaron in
long chains. \cite{pol-book}  The different dependence arises
mostly from the local anharmonic coupling in this case, and would
result in the polaron state having a higher mobility along the
chain, despite the strong coupling constant.  \\

 \noindent {\bf Conclusion}.
We have introduced the idea of librons interacting with electrons
in a complex molecular system.  The strong coupling of rotational
DOF yields a unique class of {\em twisting} polarons.  Inclusion
of anharmonic coupling to the electron via hopping and a nonlinear
restoring potential, were shown to result in rather different
polaronic behavior.  We find electronic bandwidth suppression
thanks to the excitation of twisting phonons, just as in the case
of the usual vibrational DOF, but with a comparatively weaker
effect in our case.  This difference is due to both of the
nonlinear characteristics mentioned, and provides a possible route
for differentiation in experiments.  We anticipate that the twist
polaronic features could be tested by performing experiments in a
variety of different molecules, where the restoring force is
changed systematically, for example.  We expect that our model
will provide some insights on the low-temperature transport
properties of interesting systems, such as recent molecular
electronics experiments, and helical polymer ``soft" crystals.
\c{soft} Notice however that a detailed comparison with
experiments will require one to consider the coupling of these
polarons to charge or current leads.  Such effects in vibrational
polarons have only recently been studied, \cite{N01} and one would
anticipate qualitative differences, which will be discussed
elsewhere. \\

We acknowledge support from US DOE grant no.\ DE-FG02-91ER45334,
NSF NIRT grant no.\ 0103034, and from the Condensed Matter and
Surface Sciences Program at Ohio University.  Helpful discussions
with the NIRT group at OU are also appreciated.

\begin{figure}[b]
\caption{Effective tunneling between sites, $t_{1}^{'}$, as a
function of twist mode restoring force, $K.$  $t_{1}^{'}$ is given
by Eq.\ (9) for large $K$ values, and is linear for small $K$, all
in the non-adiabatic regime. Inset: diagram of parallel twist
deformation of stack of molecular units with rotational degree of
freedom along the stack.}
\end{figure}

\includegraphics[width=14cm]{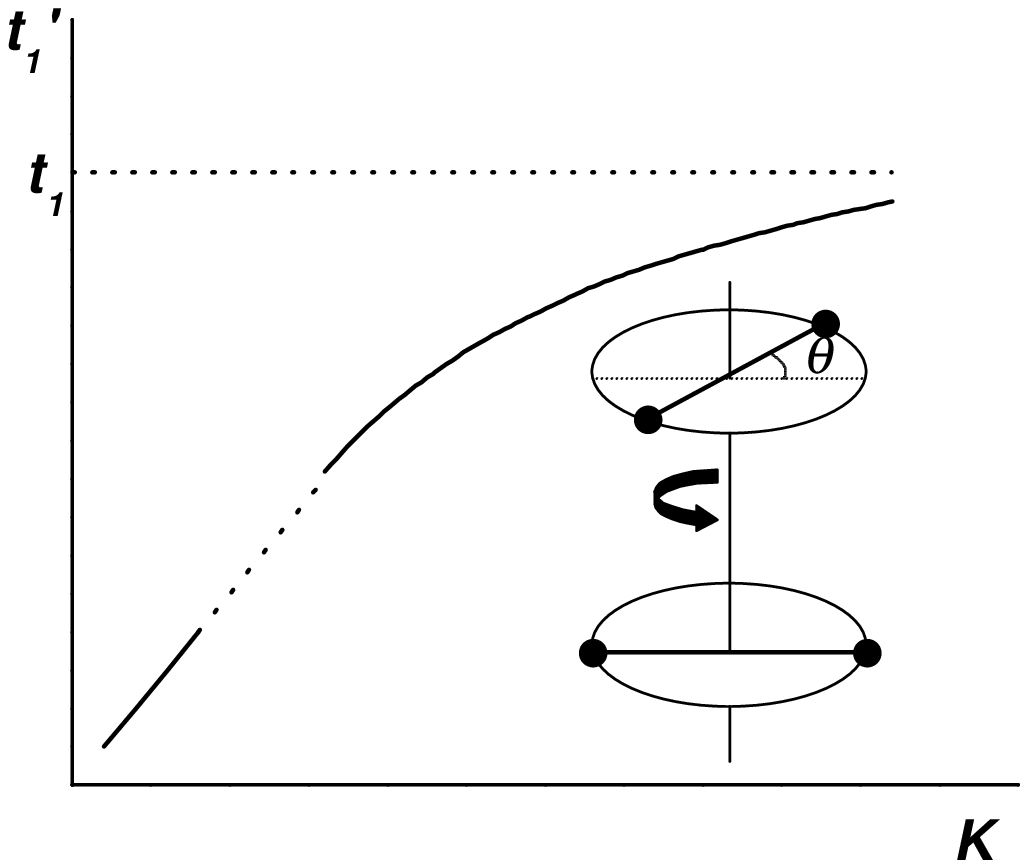}

\end{document}